\documentclass[reprint,twocolumn,aps,prc,a4paper,superscriptaddress,showpacs,preprintnumbers,amsmath,amssymb]{revtex4}

\usepackage{graphicx}
\usepackage{dcolumn}
\usepackage{bm}

\begin{document}

\title{Variation after full projection with triaxially deformed nuclear mean field}

\author{ Zao-Chun Gao}
\affiliation{China Institute of Atomic Energy, Beijing 102413, China}
\affiliation{State Key Laboratory of Theoretical Physics, Institute of Theoretical Physics, Chinese Academy of Sciences, Beijing 100190, China.}

\author{Mihai Horoi}
\affiliation{Department of Physics, Central Michigan University, Mount Pleasant, Michigan 48859, USA}

\author{ Y.S. Chen}
\affiliation{China Institute of Atomic Energy, Beijing 102413, China}

\date{\today}

\begin{abstract}
We implemented a variation after projection (VAP) algorithm based on a triaxially deformed Hartree-Fock-Bogoliubov vacuum state. This is the first projected mean field study that includes all the quantum numbers (except parity), i.e., spin ($J$), isospin ($T$) and mass number ($A$). Systematic VAP calculations with $JTA$-projection have been performed for the even-even $sd$-shell nuclei with the USDB Hamiltonian. All the VAP ground state energies are within 500 keV above the exact shell model values. Our VAP calculations show that the spin projection has two important effects: (1) the spin projection is crucial in achieving good approximation of the full shell model calculation. (2) the intrinsic shapes of the VAP wavefunctions with spin projection are always triaxial, while the Hartree-Fock-Bogoliubov methods likely provide axial intrinsic shapes. Finally, our analysis suggests that one may not be possible to associate an intrinsic shape to an exact shell model wave function.
\end{abstract}

\pacs{21.60.Jz,21.60.Cs,21.10.Hw}
\maketitle

\section{Introduction}
\label{s1}
Hartree-Fock-Bogoliubov (HFB) method has been very successful in describing the global properties of the ground states throughout the whole nuclear region. As a mean field method, HFB breaks the symmetries of the nuclear system, and can be used to study the intrinsic shapes. The HFB calculations with Gogny force show that almost all the calculated 1712 nuclei have axially symmetric HFB minima\cite{DE10}.

Projection can be done on a HFB vacuum to recover the symmetries that the Hamiltonian obeys. To test the quality of the projected wavefunctions, one can compare them with the exact shell model ones using a common Hamiltonian. HFB and variation after projected HFB calculations with shell model Hamiltonians have been reported by several authors \cite{MA11,RO08,RB11,Schmid04}. For those calculations without projection, the HFB vacuum states are often assumed to be axially symmetric \cite{RB11}.
Indeed, we will see below that all the calculated HFB minima in $sd$-shell nuclei, except $^{24}$Mg, are exactly axial with the USDB Hamiltonian \cite{usdb}.

However, if one performs the variation of the projected HFB vacuum, usually called variation after projection (VAP) \cite{Ring80}, it is likely that the intrinsic shape may changes due to the inclusion of beyond mean field correlations. One typical example is the ground state of $^{32}$Mg, which is predicted to be spherical
at the mean field level \cite{Reinhard99}, but it turns out to have a quadrupole deformation when the correlations associated
with the restoration of the broken rotational symmetry are considered \cite{Rodr00}. Another example is $^{56}$Ni, whose ground state is spherical at the mean field level, but is slightly deformed when performing the projected energy surface calculation \cite{Gao09b}.

Moreover, the triaxial ($\gamma$) degree of freedom plays important roles on the low-lying collective dynamics in this mass region \cite{Kurath72}. In $^{24}$Mg the possibility of the triaxial deformation in the ground state was discussed for decades \cite{Bonche87,Koepf88,Sheline88}, and it is still being used as the testing ground for modern theories involving angular momentum (spin) projection \cite{Bender08,Yao10,Rodr10}.

In this work, we perform  VAP calculations of the even-even $sd$-shell nuclei using the USDB Hamiltonian. Here, we allow the $\gamma$ degree of freedom in the HFB transformation. The shell model Hamiltonian conserves the spin ($J$), isospin ($T$), as well as the mass number ($A$). Hence a complete projection should recover all $J,\ T$, and $A$ quantum numbers. This is generally very much time-consuming because of the 7-dimensional integration ( 3 for $J$, 3 for $T$, and 1 for $A$). Presently, we can only carry out such extensive studies in the $sd$ shell. For efficiency, we use the new techniques of Refs. \cite{gao14,hu14,wang14} to evaluate the kernels for projections.

\section{The VAP method}

From a randomly chosen HFB vacuum state $|\Phi_0\rangle$, one can construct a new HFB vacuum state
$|\Phi\rangle$ using the Thouless theorem \cite{Ring80}. Namely,
\begin{eqnarray}\label{thouless}
|\Phi\rangle=\mathcal{N}e^{\frac12\sum_{\mu\nu}d_{\mu\nu}a^\dagger_{\mu}a^\dagger_{\nu}}|\Phi_0\rangle,
\end{eqnarray}
where $d$ is a skew symmetric matrix, and $\mathcal{N}$ is the normalization factor. The triaxiality of the HFB vacuum can be treated similar to Ref. \cite{Hara95} so that the $Q_{2\pm1}$ components of the quadrupole moment vanish.

Projecting $|\Phi\rangle$ onto good quantum numbers $J$, $T$, and $A$ one gets the so called $JTA$-projection (similarly, $TA$-projection for $T,A$, etc.). One can evaluate the $JTA$-projected energy,
\begin{eqnarray}\label{jta0}
E_{JTA}=\sum_{MKM_TK_T}f^*_{MM_T}f_{KK_T}{\langle\Phi|\hat HP^J_{MK}P^T_{M_TK_T}P^A|\Phi\rangle},
\end{eqnarray}
where $P^J_{MK}$, $P^T_{M_TK_T}$ and $P^A$ are the spin, isospin and mass number projection operators, respectively. The isospin projection operator is similar
to the spin projection operator but in the isospin space.
$E_{JTA}$ and the corresponding coefficients $f_{KK_T}$ are obtained by solving
\begin{eqnarray}\label{jta1}
\sum_{KK_T}\langle\Phi|(\hat H-E_{JTA})P^J_{MK}P^T_{M_TK_T}P^A|\Phi\rangle f_{KK_T}=0,
\end{eqnarray}
with $f_{KK_T}$ satisfying
\begin{eqnarray}\label{jta2}
\sum_{MKM_TK_T}f^*_{MM_T}f_{KK_T}\langle\Phi|P^J_{MK}P^T_{M_TK_T}P^A|\Phi\rangle=1.
\end{eqnarray}

One can also calculate $E_{TA}$ with $TA$-projection by simply removing the spin projection from Eqs. (\ref{jta0}-\ref{jta2}). For the $A$-projection, the corresponding energy, $E_A$, is reduced to
\begin{eqnarray}
E_A=\frac{\langle\Phi|\hat HP^A|\Phi\rangle}{\langle\Phi|P^A|\Phi\rangle}.
\end{eqnarray}
 Without any projection, we define
\begin{eqnarray}
E_{HFB}=\langle\Phi|\hat H|\Phi\rangle.
\end{eqnarray}

It is natural that one may consider the neutron ($N$) and proton ($Z$) projection, as has been done in Refs. \cite{MA11,Schmid04}. However, this is essentially the same as the $T_ZA$-projection ($T_Z=(N-Z)/2$). Here, we prefer to take $TA$-projection to recover the total isospin symmetry.

  VAP calculations can be performed by changing the $d$ matrix in Eq.(\ref{thouless}). Here, we impose the following restrictions for the $d$ matrix: (1) $d$ is real, (2) keeping the time reversal symmetry, and (3) no mixing between neutron and proton in the HFB transformation. Therefore the total number of free VAP parameters for $sd$-shell is reduced to $N_{VAP}=42$. In practice we start with $d=0$ and with Nilsson+BCS vacuum states
$|\Phi_0 \rangle$ obtained with randomly chosen quadrupole parameters \cite{Gao09b}.

To extract the intrinsic shape, the quadrupole moment and the triaxial degree of freedom, $Q$ and $\gamma$, are defined such that
\begin{eqnarray}\label{q}
Q\cos\gamma&=&\langle\Psi|\sqrt{\frac{16\pi}{5}}\frac{r^2}{b^2}Y_{20}|\Psi\rangle,\\
Q\sin\gamma&=&\langle\Psi|\sqrt{\frac{16\pi}{5}}\frac{r^2}{b^2}\frac{1}{\sqrt{2}}(Y_{22}+Y_{2-2})|\Psi\rangle,\label{gamma}
\end{eqnarray}
where $b$ is the harmonic oscillator length. $|\Psi\rangle$ refers to an intrinsic state, which may have different forms.
Explicitly, we define,

(1) $Q_{HFB}$ and $\gamma_{HFB}$ for $|\Psi\rangle=|\Phi\rangle$,

(2) $Q_{A}$ and $\gamma_{A}$ for  $|\Psi\rangle=\frac{P^A|\Phi\rangle}{\sqrt{\langle \Phi|P^A|\Phi\rangle}}$, and

(3) $Q_{TA}$ and $\gamma_{TA}$ for $|\Psi\rangle=\sum_{K_T}f_{K_T}P^T_{M_TK_T}P^A|\Phi\rangle$,

\section{ VAP calculations for $^{24}$Mg }

When performing the energy variation, one may find that there might be more than one energy minima. Therefore, the energy variation should be calculated several times with different starting $|\Phi_0\rangle$ states which are randomly chosen. We then identify the lowest minimum, and denote it with $E^*$. Here and below, we only discuss the results corresponding to $E^*$.

In the present work, we adopt the USDB Hamiltonian \cite{usdb}. The HFB energy for $^{24}$Mg is $E^*_{HFB}=-80.965$ MeV with the constraints $\langle \Phi |\hat N|\Phi\rangle=N$ and $\langle \Phi |\hat Z|\Phi\rangle=Z$. This is the only $sd$-shell nucleus for which the HFB calculation gives a non-axial shape with $Q^*_{HFB}=18.659$ and $\gamma^*_{HFB}=11.96^\circ$
(here and below the $Q^*$ and $\gamma^*$ are the shape parameters that can be associated with the absolute minimum for some VAP choice).
Let's first do the simplest VAP with only $A$-projection (called VAP-A). Since the particle number is already projected out, it might be unnecessary to impose a constraint to the average particle number of the HFB vacuum. To check this conjecture, we start from several different $|\Phi_0\rangle$ states and perform VAP-A. The results for few selected $\left| \Phi_0 \right>$ choices are shown in Table \ref{E_A}. One can see that the VAP-A energies are identical ($E^*_A=-81.358$ MeV). However, the corresponding $E_{HFB}$, $Q_{HFB}$, $\gamma_{HFB}$ and $\langle \hat A \rangle\equiv\langle\Phi| \hat A |\Phi\rangle$ appear randomly. This means that the converged vacua $|\Phi\rangle$ are not unique, but correspond to the same $A$-projected state. The $Q_A$ values are the same, and although the $\gamma_A$ values look different, they actually represent the same shape but with different orientations. Therefore, one can adopt the values $Q^*_A=18.284$ and $\gamma^*_A=10.05$ to define the shape of the VAP-A minimum.  If one imposes $\langle \hat A \rangle=A=8$, we still have $E^*_A=-81.358$ MeV, now the converged $|\Phi\rangle$ vacuum becomes unique, with $E_{HFB}=-79.720$, $Q_{HFB}=17.905$, and $\gamma_{HFB}=9.05^\circ$ (see the last line in Table \ref{E_A}). However, for the VAP with $TA$-projection, the situation becomes a little different.

\begin{table}
\caption{\label{E_A}Results of the VAP-A calculations for $^{24}$Mg. We perform the VAP calculations for several times. Each time we start with different $|\Phi_0\rangle$ states. The numbers in the first column denote different $|\Phi_0\rangle$ states. The second column shows the converged energy $E^*_A$. Quantities in other columns are calculated with the converged $|\Phi\rangle$ vacua. Energies are in MeV.}
\begin{ruledtabular}
\begin{tabular}{c|ccc|ccc|c}
  $|\Phi_0\rangle$ & $E^*_{A}$ & $Q_{A}$ & $\gamma_{A}(^\circ)$ & $E_{HFB}$ & $Q_{HFB}$ & $\gamma_{HFB}(^\circ)$ & $\langle \hat A \rangle$ \\
   \hline
   1&   -81.358&    18.284&    10.05&   -81.008&    18.005&     9.46&     8.110\\
   2&   -81.358&    18.284&   130.05&   -90.178&    18.371&   128.94&     9.013\\
   3&   -81.358&    18.284&  -109.95&   -82.684&    18.120&  -110.61&     8.259\\
   4&   -81.358&    18.284&    10.05&   -79.720&    17.905&     9.05&     8.000\\
\end{tabular}
\end{ruledtabular}
\end{table}

VAP calculations with $TA$-projection (called VAP-TA) are listed in Table \ref{E_TA}. Unlike VAP-A, even if one imposes $\langle \hat A \rangle=A=8$ for $^{24}$Mg,  the converged $|\Phi\rangle$ is still not unique as the $E_{HFB}$ energy appears randomly. Moreover, the $E_{A}$ energy is not unique either. Interestingly, after $TA$-projection, those different $|\Phi\rangle$ vacuum states  have exactly the same projected energy $E^*_{TA}=-82.831$(MeV) and the same $Q_{TA}=17.295$. Therefore, we can associate the shape parameter corresponding $Q^*_{TA}=Q_{TA}=17.295$ to this projected minimum. Similarly, we found (after rotation) $\gamma^*_{TA}=\gamma_{TA}=0.09^\circ$, which describes an almost axial-shape. One can conclude that only $Q_{TA}$ and $\gamma_{TA}$ are meaningful in describing the shape of the VAP-TA projected state.

\begin{table*}
\caption{\label{E_TA}Similar to Table. \ref{E_A} but for the VAP-TA calculations. $\langle A \rangle=8$ is imposed.}
\begin{ruledtabular}
\begin{tabular}{c|ccc|ccc|ccc|c}
  $|\Phi_0\rangle$ & $E^*_{TA}$(MeV) & $Q_{TA}$ & $\gamma_{TA}(^\circ)$ & $E_{A}$(MeV) & $Q_{A}$ & $\gamma_{A}(^\circ)$ & $E_{HFB}$(MeV) & $Q_{HFB}$ & $\gamma_{HFB}(^\circ)$ & $\langle A \rangle$ \\
   \hline
    1&   -82.831&    17.295&  -119.91&   -75.826&    16.376&  -118.62&   -74.921&    15.755&  -118.23&     8.000\\
    2&   -82.831&    17.295&     0.09&   -74.402&    16.167&     2.47&   -73.909&    15.563&     3.06&     8.000\\
    3&   -82.831&    17.295&   120.09&   -76.633&    16.526&   120.09&   -75.525&    15.897&   120.08&     8.000\\

\end{tabular}
\end{ruledtabular}
\end{table*}

A complete symmetry restoration is the $JTA$-projection. VAP results with $JTA$-projection (called as VAP-JTA) are shown in Table \ref{E_JTA}. All the converged $E^*_{JTA}$ energies are $-86.919$ MeV,  significantly closer to the shell model result $E_{SM}=-87.105$ MeV. Again, both $E_A$ and $E_{HFB}$ in Table \ref{E_JTA} can not be uniquely determined, even if one enforces the $\langle \hat A \rangle=A$ constraint. Fortunately, with the additional spin projection, all $E_{TA}$ values are found to be $-79.879$ MeV, and similarly the corresponding shape is described by $Q_{TA}=19.057$ and $\gamma_{TA}=16.96^\circ$. Therefore, the quantities that can be associated with the shape of VAP-JTA wavefunction should also be $Q^*_{JTA}=Q_{TA}=19.057$ and $\gamma^*_{JTA}=\gamma_{TA}=16.96^\circ$.
\begin{table*}
\caption{\label{E_JTA}Similar to Table. \ref{E_A} but for the VAP-JTA calculations. $\langle A \rangle=8$ is imposed. }
\begin{ruledtabular}
\begin{tabular}{c|c|ccc|ccc|ccc|c}
  $|\Phi_0\rangle$ & $E^*_{JTA}$(MeV) & $E_{TA}$(MeV) & $Q_{TA}$ & $\gamma_{TA}(^\circ)$ & $E_{A}$(MeV) & $Q_{A}$ & $\gamma_{A}(^\circ)$ & $E_{HFB}$(MeV) & $Q_{HFB}$ & $\gamma_{HFB}(^\circ)$ & $\langle A \rangle$ \\
   \hline
   1 &   -86.919&   -79.879&    19.057&   -16.964&   -75.600&    17.482&   -20.225&   -73.781&    16.230&   -23.772&     8.000\\
   2 &   -86.919&   -79.879&    19.057&   -16.963&   -75.641&    17.510&   -20.119&   -73.830&    16.264&   -23.604&     8.000\\
   3 &   -86.919&   -79.879&    19.057&   -16.963&   -75.644&    17.520&   -20.068&   -73.845&    16.281&   -23.506&     8.000\\
\end{tabular}
\end{ruledtabular}
\end{table*}

One can study the shape evolution of $^{24}$Mg from HFB to VAP-JTA. In VAP-TA, $Q^*_{TA}$ looks smaller than $Q^*_{HFB}$ in HFB, and $\gamma^*_{TA}$ tends to be close to zero (axial shape). However in VAP-JTA, $Q^*_{JTA}$ is larger than the $Q^*_{HFB}$ in HFB, and $\gamma^*_{JTA}$ tends to describe a triaxial shape. This triaxiality in VAP-JTA, in comparison with VAP-TA, is likely caused by the spin projection. To determine if this phenomenon is more general, we performed systematic VAP calculations for a  larger number of even-even $sd$-shell nuclei.

\section{ VAP calculations for even-even $sd$-shell nuclei }

\begin{figure}
  \includegraphics[width=3in]{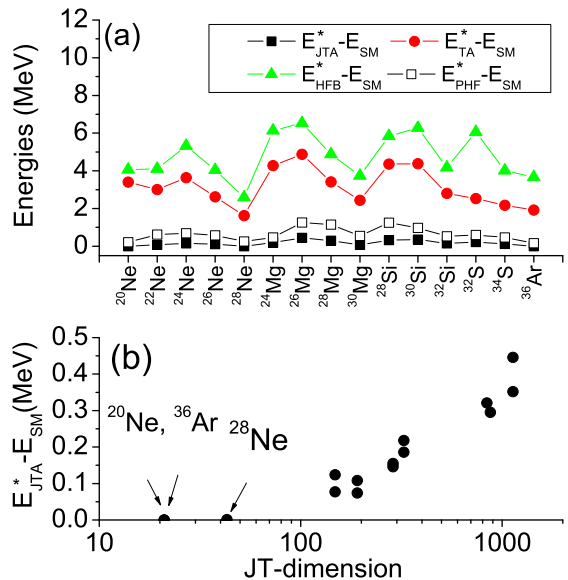}\\
  \caption{ (Color online) (a) Calculated ground state energies relative to the shell model results, $E_{SM}$. (b) Relative VAP-JTA energy, $E^*_{JTA}-E_{SM}$, versus  the shell model dimension, $N_{JT}$, in $JT$ subspace.}\label{re}
\end{figure}

VAP calculations have been performed for the ground states of even-even $sd$-shell nuclei. The calculated energies relative to the shell model ones are shown in Figure \ref{re}a.
The numerical results are given in Table \ref{sh_gs}.
Here, we didn't include the Oxygen isotopes and the $N=20$ isotones because their VAP-JTA energies are exactly the same as the shell model results ($E_{SM}$). This special case is discussed below.  The VAP-JTA energies are much lower than those of HFB and VAP-TA. Moreover, The VAP-JTA energies for $^{20}$Ne, $^{28}$Ne, and $^{36}$Ar nuclei are exactly the same as the shell model results (see also Figure \ref{re}b). This can be understood by comparing the number of VAP parameters, $N_{VAP}$, with the shell model dimension, $N_{JT}$ (the total number of the independent basis states with good $JT$). Here, $N_{VAP}=42$. The $N_{JT}$ values with $J=0$ and $T=0$ for both $^{20}$Ne and $^{36}$Ar are only $21$. For $^{28}$Ne, $N_{JT}$ for $J=0$ and $T=4$ is 43. It looks that when $N_{JT}$ is less than, or close to $N_{VAP}$, then the VAP-JTA energy is likely to be the same as the shell model one. Indeed, for all even-even oxygen isotopes and for the $N=20$ isotones, for which $N_{JT}\leq N_{VAP}$, we have obtained $E^*_{JTA}=E_{SM}$. In Figure \ref{re}b, one can also see that the energy difference $E^*_{JTA}-E_{SM}$ increases with $N_{JT}$. The largest $E^*_{JTA}-E_{SM}=0.446$ MeV is obtained for $^{26}$Mg, corresponding to the largest $N_{JT}=1132$.

\begin{table*}
\caption{\label{sh_gs} Converged energies and associated shape parameters for even-even $sd$-shell nuclei calculated  with the USDB Hamiltonian.}\begin{ruledtabular}
\begin{tabular}{ccc|ccc|ccc|ccc|ccc}
 Nucleus&&&	\multicolumn{3}{c|}{VAP-JTA}&\multicolumn{3}{c|}{VAP-TA}&\multicolumn{3}{c|}{HFB}&\multicolumn{3}{c}{VAP-HF}\\
 \hline
         &$N_{JT}$&$E_{SM}$&$E^*_{JTA}$&$Q^*_{JTA}$&$\gamma^*_{JTA}$&$E^*_{TA}$& $Q^*_{TA}$& $\gamma^*_{TA}$&$E^*_{HFB}$& $Q^*_{HFB}$& $\gamma^*_{HFB}$&$E^*_{PHF}$& $Q^*_{PHF}$& $\gamma^*_{PHF}$\\
         \hline
$^{20}$Ne&21  & -40.472& -40.472& --  &	--  & -37.069&	14.7&  0.0&	 -36.404&	15.3&	 0.0&  -40.265&13.861& 3.551\\
$^{22}$Ne&148 & -57.578& -57.501&  12.1&	13.8& -54.572&	15.8&	 0.0&	 -53.474&	16.5&	 0.0&  -56.958&15.675& 8.632\\
$^{24}$Ne&287 & -71.725& -71.570&  11.0&	30.1& -68.084&	10.1&	60.0&	 -66.402&	12.0&	 0.0&  -71.037&13.449&32.786\\
$^{26}$Ne&191 & -81.564& -81.465&  9.2 &	28.4& -78.949&	 8.6&	 0.0&	 -77.518&	 8.3&	 0.0&  -80.988& 9.760&17.265\\
$^{28}$Ne&43  & -86.543& -86.543&	--  &  -- & -84.920&	 7.0&	60.0&	 -83.949&	 7.1&	 0.0&  -86.294& 9.848&23.934\\
$^{24}$Mg&325 & -87.105& -86.919&	19.1&	17.0& -82.831&	17.3&	 0.0&	 -80.965&	18.7&	12.0&  -86.636&19.165&16.427\\
$^{26}$Mg&1132&-105.521&-105.075&	15.8&	28.7&-100.648&	13.8&	25.7&	 -98.992&	15.9&	60.0& -104.264&16.238&32.331\\
$^{28}$Mg&874 &-120.500&-120.205&	14.5&	20.2&-117.091&	14.4&	 0.0&	-115.625&	15.1&	 0.0& -119.354&16.306&19.835\\
$^{30}$Mg&191 &-130.474&-130.400&	10.4&	20.0&-128.035&	10.3&	 0.0&	-126.735&	10.9&	 0.0& -129.926&11.864&27.322\\
$^{28}$Si&839 &-135.860&-135.539&	16.1&	58.6&-131.501&	17.8&	60.0&	-130.021&	19.8&	60.0& -134.617&17.116&58.038\\
$^{30}$Si&1132&-154.754&-154.402&	14.3&	47.0&-150.380&	10.6&	60.0&	-148.475&	14.5&	60.0& -153.777&14.633&46.167\\
$^{32}$Si&287 &-170.519&-170.373&	12.5&	58.2&-167.721&	10.6&	60.0&	-166.344&	12.4&	60.0& -169.996&12.175&52.023\\
$^{32}$S &325 &-182.452&-182.234&	15.1&	33.3&-179.925&	 0.6&	60.0&	-176.393&	 0.0&	 0.0& -181.856&14.681&33.188\\
$^{34}$S &148 &-202.504&-202.380&	10.6&	53.0&-200.331&	 0.0&	 0.0&	-198.493&	 0.0&	 0.0& -202.039&11.496&50.992\\
$^{36}$Ar&21  &-230.277&-230.277&	 -- &  -- &-228.355&   0.0&	 0.0&	-226.611&	13.2&	60.0& -230.112&12.068&52.293
\end{tabular}
\end{ruledtabular}
\end{table*}

\begin{figure}
  \includegraphics[width=3in]{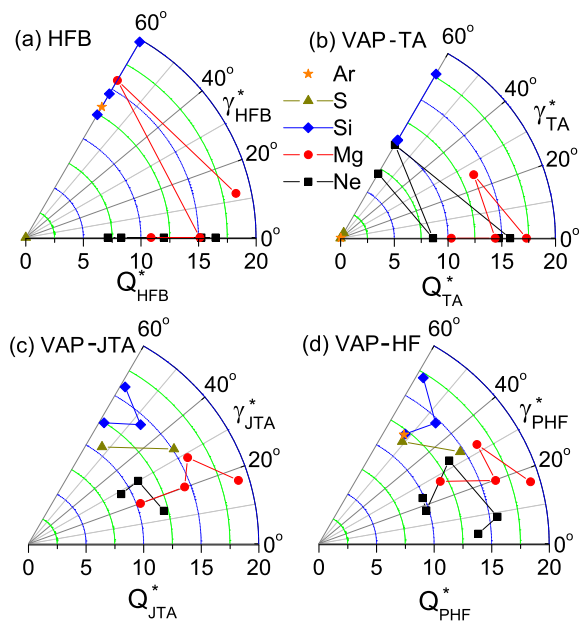}\\
  \caption{(Color online) Quadrupole moments $Q$ and $\gamma$ values for (a) HFB vacuum states, (b) VAP-TA states, (c) VAP-JTA states and (d) VAP-HF states which is based on a slater determinant. }\label{shape}
\end{figure}

The quadrupole moment and the $\gamma$ degree of freedom can be extracted using Eqs. (\ref{q}) and (\ref{gamma}). In Fig. \ref{shape}a, the $\gamma^*_{HFB}$ values in HFB are either $0^\circ$ or $60^\circ$, except $\gamma^*_{HFB}=12^\circ$ for $^{24}$Mg, thus supporting the conclusion that HFB likely presents axially deformed shapes. In Fig. \ref{shape}b, the shapes in VAP-TA calculations still remain axially symmetric, except for $^{26}$Mg, which has $\gamma^*_{TA}=25.7^\circ$. Quite differently, the $\gamma^*_{JTA}$ values in the VAP-JTA calculations (Fig. \ref{shape}c) show that all these nuclei are non-axial without exception. Comparing these results with those of  Fig. \ref{shape}a, one can conclude that the triaxiality in VAP-JTA is definitely a beyond mean-field effect, which is likely to be a universal phenomenon. Fig. \ref{shape}b, however, excludes the possibility that the isospin projection and the mass projection lead to triaxiality. Thus, the only possible cause of the triaxiality is the beyond mean-field spin projection.

To study directly the effect of spin projection, one can start from a Hartree-Fock (HF) Slater determinant (SD) and perform VAP calculations with only spin projection (called VAP-HF). The converged energies, $E^*_{PHF}$, relative to $E_{SM}$, are shown in Fig. \ref{re}a. The results show that VAP-HF is better than VAP-TA, and quite close to the VAP-JTA. The quadrupole moment $Q^*_{PHF}$ and $\gamma^*_{PHF}$ corresponding to $E^*_{PHF}$ can be calculated using Eqs. (\ref{q}) and (\ref{gamma}) with $|\Psi\rangle$ replaced by the converged SD. These quantities are uniquely determined, and are shown in Fig. \ref{shape}d. Again, all the $\gamma^*_{PHF}$ values are distributed in the interval $(0^\circ, 60^\circ)$, which is very similar to Fig. \ref{shape}c. Therefore, we could conclude that VAP results that include spin projection can always be associated with intrinsic states having triaxial deformation.

One more interesting phenomenon, however, is related to the VAP-JTA calculations for $^{20}$Ne, $^{28}$Ne, and $^{36}$Ar. We have shown above that the $E^*_{JTA}$ energies of these nuclei are the same as the exact shell model results. Surprisingly, the corresponding $Q_{TA}$ and $\gamma_{TA}$ values are not unique, which is quite different from other nuclei with $E^*_{JTA}>E_{SM}$. For example, the results for $^{20}$Ne are shown in Table \ref{ne20}. With the same converged $E^*_{JTA}=-40.472$MeV, one can clearly see that starting with different initial states $|\Phi_0\rangle$, the result for $Q_{TA}$ and $\gamma_{TA}$ could be different. These results indicate that it may not be possible to associate an unique intrinsic deformation with an exact eigenstate of the Hamiltonian.

\begin{table}
\caption{\label{ne20}VAP results with JTA projection for $^{20}$Ne.}
\begin{ruledtabular}
\begin{tabular}{c|c|ccc}
  $|\Phi_0\rangle$ & $E^*_{JTA}$(MeV) & $E_{TA}$(MeV) & $Q_{TA}$ & $\gamma_{TA}(^\circ)$ \\
   \hline
   1&    -40.472&   -28.284&     6.314&   -45.134\\
   2&    -40.472&   -30.468&    11.873&  -124.746\\
   3&    -40.472&   -27.932&     9.876&     2.592\\
\end{tabular}
\end{ruledtabular}
\end{table}

\section{ Summary }
We implemented an algorithm that performs variation after projection (VAP) on spin, isospin, and mass number  of a triaxially deformed Hartree-Fock-Bogoliubov vacuum state. This is the first projected mean field study that includes all these quantum numbers.

We start from a randomly chosen HFB vacuum state and carry out VAP calculations for $^{24}$Mg in $sd$-shell with various projections. In the VAP-A case the converged solution is independent of the Fermi level (chemical potential). Although the associated HFB vacuum does not have definite quadrupole moment $Q_{HFB}$ and triaxial deformation parameter $\gamma_{HFB}$, one can use the unique $Q_A$ and $\gamma_A$ to describe the intrinsic deformation of the VAP-A state. Similarly, in the VAP-TA calculations, $Q_A$ and $\gamma_A$ can not be uniquely determined, but  $Q_{TA}$ and $\gamma_{TA}$ are unique and can be associated with the intrinsic deformation of the VAP-TA state. It is not possible to directly define deformation parameters $Q$ and $\gamma$ for the VAP-JTA wave function, which has the symmetries fully restored, but the $Q_{TA}$ and $\gamma_{TA}$ calculated with the VAP-JTA vacuum state $|\Phi \rangle$ are also unique, and can be associated with the intrinsic deformation of the VAP-JTA state.

Systematical VAP calculations of even-even $sd$-shell nuclei have been performed using the USDB Hamiltonian. The VAP-JTA energies, $E^*_{JTA}$, are very close to the shell model results, $E_{SM}$. Moreover, the relative energy, $E^*_{JTA}-E_{SM}$, increases with the shell model dimension $N_{JT}$. The shapes described by the HFB minima are always axial. However, with spin projection VAP calculations always produce triaxial shapes. We believe that such triaxiality is an universal phenomenon caused by the beyond mean-field dynamic correlations. Finally, we show that those VAP-JTA states reaching the exact shell model results do not have clearly defined intrinsic shapes.

\begin{acknowledgments}
Z-C. Gao thanks Prof. G. F. Bertsch for fruitful discussions, and Prof. Y. Sun for reading the manuscript. This work is partly supported by the
National Natural Science Foundation of China under Contract No.s
11175258, 11575290, 11321064 and 11275068; the CUSTIPEN
(China-U.S. Theory Institute for Physics with Exotic Nuclei)
under DOE Grant No. DE-FG02-13ER42025; the Open Project Program of State Key Laboratory of Theoretical Physics, Institute of Theoretical Physics, Chinese Academy of Sciences, China (No.Y5KF141CJ1). M.H. acknowledges the U.S. NSF Grant No. PHY-1404442.
\end{acknowledgments}


%


\begin{thebibliography}{10}
\bibitem{DE10}J.-P. Delaroche et al., Phys. Rev. C 81, 014303 (2010).
\bibitem{MA11} I. Maqbool, J. A. Sheikh, P. A. Ganai and P. Ring, J. Phys. G{\bf 38} 045101 (2011).
\bibitem{RO08}R. Rodr\'{\i}guez-Guzm\'{a}n, Y. Alhassid, and G. F. Bertsch, Phys.
Rev C 77, 064308 (2008).
\bibitem{RB11}L. M. Robledo and G. F. Bertsch, Phys. Rev. C 84, 014312 (2011).
\bibitem{Schmid04}K.W. Schmid, Prog. Part. Nucl. Phys. {\bf 52}, 565 (2004).
\bibitem{usdb}B. A. Brown and W. A. Richter, Phys. Rev. C 74, 034315 (2006).
\bibitem{Ring80}P. Ring and P. Schuck, The Nuclear Many-Body Problem
(Springer Verlag, New York, Heidelberg, Berlin, 1980).
\bibitem{Reinhard99} P.-G. Reinhard, D. J. Dean, W. Nazarewicz, J. Dobaczewski,
J. A. Maruhn, and M. R. Strayer, Phys. Rev. C 60, 014316
(1999).
\bibitem{Rodr00}R. R. Rodr\'{\i}guez-Guzm\'{a}n, J. L. Egido, and L. M. Robledo, Phys.
Lett. B474, 15 (2000); Phys. Rev. C 62, 054319 (2000).
\bibitem{Gao09b}Zao-Chun Gao, Mihai Horoi, and Y. S. Chen, Phys. Rev. C{\bf 80}, 034325 (2009).
\bibitem{Kurath72} D. Kurath, Phys. Rev. C 5, 768 (1972).
\bibitem{Koepf88} W. Koepf and P. Ring, Phys. Lett. B 212, 397 (1988).
\bibitem{Bonche87} P. Bonche, H. Flocard, and P. H. Heenen, Nucl. Phys. A 467,
115 (1987).
\bibitem{Sheline88}R. K. Sheline, I. Ragnarsson, S. {\AA}berg, and A.Watts, J. Phys. G
14, 1201 (1988).
\bibitem{Bender08} Michael Bender and Paul-Henri Heenen Phys. Rev. C{\bf 78}, 024309 (2008).
\bibitem{Yao10}J. M. Yao, J. Meng, P. Ring, and D. Vretenar Phys. Rev. C{\bf 81}, 044311 (2010).
\bibitem{Rodr10}Tom\'{a}s R. Rodr\'{\i}guez and J. L. Egido Phys. Rev. C{\bf 81}, 064323 (2010).
\bibitem{gao14}Z.-C. Gao, Q.-L. Hu, Y.S.Chen Phys. Lett.  B732, 360 (2014).
\bibitem{hu14}Q.-L. Hu, Z.-C. Gao, Y.S.Chen Phys. Lett.  B734, 162 (2014).
\bibitem{wang14}L.-J. Wang, F.-Q. Chen, T. Mizusaki, M. Oi, Y. Sun, Phys. Rev. C 90 (2014) 011303(R)
\bibitem{Hara95} K. Hara, Y. Sun, Int. J. Mod. Phys. E 4, 637 (1995).

\end{thebibliography}
\end{document}